\documentclass[twocolumn]{aastex631}

\usepackage{xspace,xcolor,ulem}

\graphicspath{{./}{Fig/}}
\graphicspath{{Fig/}}

\newcommand{\TWAB}{TWA\,27\,B\xspace}

\newcommand{\PDSbc}{PDS\,70\,b and c\xspace}
\newcommand{\Delormeb}{Delorme\,1\,(AB)\,b\xspace}
\newcommand{\NIRSpec}{JWST/NIRSpec}
\newcommand{\Lya}{{\ifmmode \text{Ly-}\alpha \else Ly-$\alpha$\fi}\xspace}
\newcommand{\Pa}{{\ifmmode \text{Pa-}\alpha \else Pa-$\alpha$\fi}\xspace}
\newcommand{\Pb}{{\ifmmode \text{Pa-}\beta \else Pa-$\beta$\fi}\xspace}
\newcommand{\Pg}{{\ifmmode \text{Pa-}\gamma \else Pa-$\gamma$\fi}\xspace}
\newcommand{\Pd}{{\ifmmode \text{Pa-}\delta \else Pa-$\delta$\fi}\xspace}

\newcommand{\Bra}{{\ifmmode \text{Br-}\alpha \else Br-$\alpha$\fi}\xspace}
\newcommand{\Brb}{{\ifmmode \text{Br-}\beta \else Br-$\beta$\fi}\xspace}
\newcommand{\Brg}{{\ifmmode \text{Br-}\gamma \else Br-$\gamma$\fi}\xspace}
\newcommand{\Brd}{{\ifmmode \text{Br-}\delta \else Br-$\delta$\fi}\xspace}

\newcommand{\Ha}{{\ifmmode \text{H}\alpha \else H$\alpha$\fi}\xspace}
\newcommand{\Hb}{{\ifmmode \text{H}\beta \else H$\beta$\fi}\xspace}

\newcommand{\Ftwo}{240pt}
\newcommand{\MP}{M_\mathrm{P}}
\newcommand{\RP}{R_\mathrm{P}}

\newcommand{\MJ}{M_\mathrm{J}}
\newcommand{\RJ}{R_\mathrm{J}}
\newcommand{\Mdot}{\dot{M}}

\newcommand{\AV}{A_\mathrm{V}}
\newcommand{\vff}{v_\mathrm{ff}}
\newcommand{\ff}{\ensuremath{f_\mathrm{f}}\xspace}
\newcommand{\lami}{\lambda_i}

\newcommand{\vshift}{\ensuremath{v_\mathrm{shift}}\xspace}
\newcommand{\vcrit}{\ensuremath{v_\mathrm{crit}}\xspace}
\newcommand{\rcrit}{\ensuremath{r_\mathrm{crit}}\xspace}
\newcommand{\np}{n_\mathrm{p}}
\newcommand{\nAF}{n_\mathrm{AF}}
\newcommand{\TAF}{T_\mathrm{AF}}
\newcommand{\kms}{{ \ifmmode \mathrm{km\,s^{-1}} \else km\,s$^{-1}$\fi}}
\newcommand{\cc}{{ \ifmmode \mathrm{cm^{-3}} \else cm$^{-3}$\fi}\xspace}
\newcommand{\MJyr}{{ \ifmmode \mathrm{\MJ\,yr^{-1}} \else $\MJ$\,yr$^{-1}$\fi}}
\newcommand{\redchi}{ \chi^2_\mathrm{red}}
\newcommand{\flam}{f_\lambda}
\newcommand{\Fobs}{F_\mathrm{obs}}
\newcommand{\Fmod}{F_\mathrm{mod}}
\newcommand{\Fmock}{F_\mathrm{mock}}

\newcommand{\revise}[1]{{#1}\xspace}

\begin{document}

\title{Analyzing JWST/NIRSpec Hydrogen Line Detections at TWA 27B: Constraining Accretion Properties and Geometry}
\shorttitle{Analyzing Hydrogen Lines from TWA 27B}
\shortauthors{Aoyama et al.}

\author[0000-0003-0568-9225]{Yuhiko Aoyama}
\affiliation{
Kavli Institute for Astronomy and Astrophysics, Peking University, Beijing 100084, People’s Republic of China
}
\affiliation{
School of Physics and Astronomy, Sun Yat-sen University, Guangdong 519082, People's Republic of China
}
\correspondingauthor{Yuhiko Aoyama}
\email{yaoyama@pku.edu.cn}

\author[0000-0002-2919-7500]{Gabriel-Dominique Marleau}
\affiliation{%
Max-Planck-Institut f\"ur Astronomie,
K\"onigstuhl 17,
69117 Heidelberg, Germany
}

\affiliation{
Fakult\"at f\"ur Physik,
Universit\"at Duisburg-Essen,
Lotharstra\ss{}e 1,
47057 Duisburg, Germany
}
\affiliation{%
Institut f\"ur Astronomie und Astrophysik,
Universit\"at T\"ubingen,
Auf der Morgenstelle 10,
72076 T\"ubingen, Germany
}
% \affiliation{%
% Physikalisches Institut,
% Universit\"{a}t Bern,
% Gesellschaftsstr.~6,
% 3012 Bern, Switzerland}

\author[0000-0002-3053-3575]{Jun Hashimoto}
\affil{Astrobiology Center, National Institutes of Natural Sciences, 2-21-1 Osawa, Mitaka, Tokyo 181-8588, Japan}
\affil{Subaru Telescope, National Astronomical Observatory of Japan, Mitaka, Tokyo 181-8588, Japan}
\affil{Department of Astronomy, School of Science, Graduate University for Advanced Studies (SOKENDAI), Mitaka, Tokyo 181-8588, Japan}

\begin{abstract}
Hydrogen lines from forming planets are crucial for understanding planet formation. However, the number of planetary hydrogen line detections is still limited. 
Recent JWST/NIRSpec observations have detected Paschen and Brackett hydrogen lines at TWA 27 B (2M1207b). 
TWA 27 B is classified as a planetary-mass companison (PMC) rather than a planet due to its large mass ratio to the central object ($\approx 5 M_\mathrm{J}$ compared to $25 M_\mathrm{J}$). Nevertheless, TWA 27 B's hydrogen line emission is expected to be same as for planets, given its small mass.
We aim to constrain the accretion properties and accretion geometry of TWA 27 B, contributing to our understanding of hydrogen line emission mechanism common to both PMCs and planets.
We conduct spectral fitting of four bright hydrogen lines (Pa-$\alpha$, Pa-$\beta$, Pa-$\gamma$, Pa-$\delta$) with an accretion-shock emission model tailored for forming planets. 
We estimate the mass accretion rate at $\dot{M} \approx 3 \times 10^{-9}\, M_\mathrm{J}\,\mathrm{yr}^{-1}$ with our fiducial parameters, though this is subject to an uncertainty of up to factor of ten. 
Our analysis also indicates a dense accretion flow, $n\gtrsim 10^{13}\,\mathrm{cm^{-3}}$ just before the shock, implying a small accretion-shock filling factor $f_\mathrm{f}$ on the planetary surface ($f_\mathrm{f} \lesssim 5\times10^{-4}$). This finding suggests that magnetospheric accretion is occurring at TWA 27 B.
Additionally, we carry out a comparative analysis of hydrogen-line emission color to identify the emission mechanism, but the associated uncertainties proved too large for definitive conclusions.
This underscores the need for further high-precision observational studies to elucidate these emission mechanisms fully.
\end{abstract}


\section{Introduction}
Observations of growing planets provide crucial constraints on planet formation. The process of growth through mass accretion is inherently connected to the release of gravitational energy, resulting in the heating of gas. This, in turn, leads to characteristic line emissions.
In particular, hydrogen lines are widely used to confirm and characterize the accreting objects due to their brightness \citep{Hartmann.etal2016AccretionPreMainSequenceStars}.
Indeed, a hydrogen line, \Ha, has been detected in the accreting planets \PDSbc\ \citep{Wagner.etal2018MagellanAdaptiveOptics,Haffert.etal2019Twoaccretingprotoplanets,Zhou.etal2021HubbleSpaceTelescope}, offering insights into planet formation.

Hydrogen line emission from accreting stars has been extensively studied. In the context of accreting young stars, hydrogen lines originate in the accretion flow \citep[see the review of][]{Hartmann.etal2016AccretionPreMainSequenceStars} that occurs within the magnetospheric accretion process \citep{Koenigl1991}. 
While the accretion flow generates a strong shock on the stellar surface, observed line profiles support that the emission from the preshock accretion flow rather than such postshock region \citep{Hartmann+1994,Muzerolle.etal1998MagnetosphericAccretionModels,Muzerolle.etal2001EmissionLineDiagnostics}, potentially because the postshock gas is hot enough to fully ionize atomic hydrogen, resulting in minimal hydrogen line emission.

Similarly to the accreting stars, magnetospheric accretion can occur during the accretion from the circum-planetary disks (CPDs) to planets \citep{Zhu2015ACCRETINGCIRCUMPLANETARYDISKS,Batygin2018TerminalRotationRates,Hasegawa+2021,Hasegawa+2024}. 
As for accreting stars, the preshock accretion flow can emit hydrogen lines \citep{Thanathibodee.etal2019MagnetosphericAccretionSource}. 
However, a distinct characteristic in planetary accretion is the cooler postshock gas resulting from the lower mass of planets. 
This cooler postshock condition facilitates hydrogen line emission \citep{Aoyama.Ikoma2019ConstrainingPlanetaryGas,Aoyama+2020} from the geometrically thin, non-equilibrium hot layer (known as the Zel'dovich spike in the context of radiative shocks; e.g., \citealp{Vaytet.etal2013influencefrequencydependentradiative}; see also the schematic in Fig.~2 of \citealp{Marleau.etal2022AccretingprotoplanetsSpectral}).
These preshock and postshock emissions are dominant in less and more energetic accretion cases, respectively \citep{Aoyama.etal2021ComparisonPlanetaryHaemission}. 
The accretion rate or object mass at which the dominant emission source changes depends on the preshock accretion flow temperature, which is still unclear \citep[e.g.,][]{Muzerolle.etal2001EmissionLineDiagnostics,Lima.2010ModelingHalpha}.

In scenario lacking magnetospheric accretion, accreting planets may still generate strong shock on their surface and thus signnificant hydrogen line emissions through another mechanism.
As forming planets and their CPDs receive vertical gas inflows from the surrounding protoplanetary disk, weak shocks on the CPD surface create a fast surface-accretion layer \citep{Szulagyi.etal2016Circumplanetarydisccircumplanetary,Takasao.etal2021HydrodynamicModelHa}. This accretion flow impacts the planetary surface at sub-free-fall velocity, leading to hydrogen line emission \citep{Takasao.etal2021HydrodynamicModelHa}.
However, \citet{Marleau+2023} proposed that these emissions could be negligible if the cooling after the CPD-surface shock is efficient, as predicted when using equilibrium-abundance gas opacity \citep[e.g., from][]{Malygin+2014}.
On the other hand, \citet{Takasao.etal2021HydrodynamicModelHa} estimated in their Appendix~B that the cooling timescale is determined by the CO re-formation timescale when the postshock temperature exceeds $\approx 6500$\,K and the number density is below $\approx 10^{12}$ \cc. Although the CPD-surface shock achieves these condition in the vicinity of the planet, their ignored effects, such as mixing the post-shock layer with the CPD or the abundant dust taken with the accretion flow, potentially bypasses the need for the CO re-formation and cool the postshock layer quickly.
Consequently, the presence of such fast accretion layer and its resulting emissions remains not settled.
Given the diverse range of possible accretion geometries and the emission mechanisms, elucidating the emission mechanisms requires further observational studies of planetary hydrogen lines.

After the first solid detection of planetary hydrogen lines at \PDSbc, subsequent searches for accreting planets resulted in several tens of non-detections \citep{Cugno+2019,Xie.etal2020SearchingprotoplanetsMUSE,Zurlo.etal2020widestsurveyaccreting,Huelamo+2022,Follette+2023,Cugno.etal2023MagAOXHSThighcontrast}.
The tentative detection of AB Aur b \citep{Currie.etal2022ImagesembeddedJovian} remains under debate \citep{Zhou+2022,Zhou+2023}.
Even at \PDSbc, detections have been limited to \Ha\ and recombination Balmer continuum \citep{Zhou.etal2021HubbleSpaceTelescope}, with attempts to detect other hydrogen lines proving unsuccessful \citep{Hashimoto.etal2020AccretionPropertiesPDS,Uyama.etal2017ConstrainingAccretionSignatures,Uyama.etal2021KeckOSIRISPav}.

Expanding the scope beyond planets, there are several reports of hydrogen line detections from planetary mass companions (PMCs). These PMCs, however, are not classified as of `planets' due to differences in their expected origins or forming histories.
For instance, \Delormeb, which was detected in hydrogen lines, exhibits a mass ratio to its central binaries of $\sim0.08$, suggesting a binary-like rather than a planet-like formation \citep[e.g.,][]{Lodato+2005,Reggiani+2016}. 
Nevertheless, the mechanism of hydrogen-line emission due to mass accretion depends solely on the free-fall velocity of the objects, regardless of its formation history \citep{Aoyama.etal2021ComparisonPlanetaryHaemission}. Consequently, \Delormeb\ serves as an excellent subject for studying accretion properties at planetary-mass objects through hydrogen lines \citep{Eriksson.etal2020Strongemissionsigns,Ringqvist.etal2023ResolvednearUVhydrogen}. Notably, \citet{Betti.etal2022NearinfraredAccretionSignatures,betti22c} conducted a comparative analysis of stellar-like preshock emission and planetary-like postshock emission using hydrogen-line emission colors, concluding that the postshock emission is more probable for \Delormeb. This finding strongly supports the notion that PMCs and planets undergo similar mass accretion processes and resulting hydrogen line emission, despite their different early formation histories and system architectures.

Another recent notable detection of hydrogen lines from PMCs is at \TWAB\ (2M1207b), which is also the first directly-imaged planetary-mass object \citep{Chauvin.etal2004directimaging}.
\citet{Luhman.etal2023JWSTNIRSpecObservations} reported medium-resolution ($R\approx$1700--4000) observations with \NIRSpec, with further hydrogen-line-focused analysis by \citet{Marleau+24}. \TWAB\ is an $\approx 5\,\MJ$ planet orbiting at $50.3\pm0.3$\,au, located $64.5\pm0.4$\,pc away \citep{Bailer-Jones+2021}. While the mass ratio to the central body is high ($\sim0.2$), its planetary mass, being lower than that of \Delormeb, suggests a likelihood of planet-like postshock hydrogen-line emission. Moreover, this represents the first detection of hydrogen lines from PMCs using \NIRSpec. Given the high sensitivity of JWST, \NIRSpec\ is anticipated to find more hydrogen lines from PMCs and planets in the near future. Therefore, this observation serves an excellent opportunity for exploring what can be learned about the accretion from the \NIRSpec\ hydrogen line observations.

In this paper, we conduct a detailed analysis of the \NIRSpec\ hydrogen-line data for \TWAB, proceeding under the assumption that these emission originates from the accretion-shock-heated gas. We aim to study the accretion process of this specific object and, in addition, to evaluate the extent to which \NIRSpec\ observations of planetary hydrogen lines can constrain planetary accretion properties.
We perform spectral fitting for the detected bright hydrogen lines, as detailed in Section~\ref{sec:HL_method}, and provide constraints on the accretion properties in Section~\ref{sec:HL_results}. 
In Section~\ref{sec:HL_Discussion}, we discuss the accretion geometry, the possibility of stellar-like accretion-flow emission, and implication for future observations. 
Finally, we summarize and conclude this study in Section~\ref{sec:HL_summary}.

\section{Method}
\label{sec:HL_method}
\subsection{Overview of the line emission model}
\label{sec:Emodel}
In this study, we assume the observed hydrogen lines originate from the shock-heated gas due to the accretion flow to the planet.
To model this postshock gas and its hydrogen-line emission, we use the model of \citet{Aoyama.etal2018TheoreticalModelHydrogen}. This model solves radiative transfer equations, taking into account the temporal evolution of non-equilibrium chemical reactions such as the electron transitions of atomic hydrogen. 
Effectively, the only direct input parameters for this model are the preshock velocity and hydrogen nuclei number density $(v_0, n_0)$, and the outputs are spectrally-resolved hydrogen-line fluxes from unit surface area. Due to the strong shock and consequently high temperatures, other preshock properties like temperature, chemical compositions, and external radiation fields have minimal impact on the model outputs \citep{Aoyama+2020}.

The shock model in \citet{Aoyama.etal2018TheoreticalModelHydrogen} simulates only the postshock region, focusing on the microphysical processes. It is thus independent of the broader flow geometry and applicable to any strong shock around the planet.
Specifically, it is relevant to scenarios encompassing the shock on the planetary surface via magnetospheric accretion \citep{Aoyama.Ikoma2019ConstrainingPlanetaryGas,Aoyama+2020,Aoyama.etal2021ComparisonPlanetaryHaemission}, via CPD-surface accretion layers \citep{Takasao.etal2021HydrodynamicModelHa}, or via direct infall from the PPD \citep{Marleau+2023}, as well as to the CPD surface shock \citep{Aoyama.etal2018TheoreticalModelHydrogen, Marleau+2023}.
In applying to \TWAB, we assume that a single set of input parameters ($v_0, n_0$) adequately represents the emission-source shock, which is appropriate for the planetary surface shock. 
This assumption is inappropriate for the CPD surface shock, where $v_0$ depends on the distance from the planet. However, this geometry turns out to be unlikely as discussed in \S~\ref{sec:geometry}. 
In this study, we assume all the observed hydrogen lines emerge from the postshock region, while the preshock gas can also contribute \citep{Thanathibodee.etal2019MagnetosphericAccretionSource,Aoyama.etal2021ComparisonPlanetaryHaemission}. The possible contribution from the preshock region will be discussed in \S~\ref{sec:EMechanism}.

\subsection{Fitting procedures for the modeled and observed spectra}
\label{sec:M_fit}
We fit the modeled spectra to the observed spectra, at each shock-parameter set of ($v_0, n_0$) to estimate likelihood of each shock-parameter set. We use four bright and simultaneously-observed lines (\Pa, \Pb, \Pg, and \Pd) with the G140H/F100LP grating--filter combination. 
We formulate the mock-observational flux energy density as
\begin{equation}
    \Fmock(\lambda) =  \frac{S}{ 4\pi d^2} 10^{-0.4A(\flam \lambda)}\Fmod(\flam \lambda),
    \label{eq:mock_general}
\end{equation}
where $\Fmod$ is the hydrogen-line flux per unit emitting area modeled as described in \S~\ref{sec:Emodel}, $S$ is the emitting area, $d$ is the distance to the object, $A$ is the extinction magnitude at each wavelength, and $\flam$ is the wavelength-shift factor. Since the JWST operates in space, there is no atmospheric refraction. Consequently, $\flam$ is predominantly attributed to the Doppler effect resulting from the motion of the emission source. Thus, $\flam =1 + \vshift c^{-1}$ where $\vshift$ is the radial component of the source velocity and $c$ is the speed of light. 
In this study, we assume that there is no extinction ($A=0$) and that a fraction $\ff$ of planetary surface is covered by the accretion shock, so that $S=\ff 4 \pi \RP^2$, where $\RP$ is the planetary radius.
Then, Eq.~(\ref{eq:mock_general}) can be rewritten as
\begin{equation}
    \label{eq:mock_used}
    \Fmock(\lambda) = \ff \left( \frac{\RP}{d} \right)^2 \Fmod\left( \lambda \left[1+\frac{\vshift}{c}\right] \right).
\end{equation}

Additionally, we align the modeled spectral profiles with the observed data by first convolving them with the instrumental broadening function and then average them within the width of the wavelength bins corresponding to the observational ones.
We assume the instrumental broadening function to be a Gaussian profile with Full-Width at Half-Maximum $=\lambda_0 R^{-1}$, where $\lambda_0$ is the line-center wavelength and $R$ is the instrumental spectral resolution. The specific $\lambda_0$ and $R$ values are provided in Table~\ref{tab:RR}. 

\begin{deluxetable}{ccc}
    \tablecaption{Instrumental spectral resolution \label{tab:RR}}    
    \tablecolumns{3}
    \tablehead{
    \colhead{Line} & 
    \colhead{Vacuum rest wavelength $\lambda_0$ [\AA]} & 
    \colhead{Spectral resolution $R^{[1]}$}
    }
    \startdata
        \Pa & 18756 & 3769 \\
        \Pb & 12822 & 2461 \\
        \Pg & 10941 & 2084 \\
        \Pd & 10052 & 1908 
    \enddata
    \tablecomments{[1] G140H/F100LP grating and filter, from the ``NIRSpec Dispersers and Filters'' page (\url{https://jwst-docs.stsci.edu/jwst-near-infrared-spectrograph/nirspec-instrumentation/nirspec-dispersers-and-filters}). }
\end{deluxetable}

We fit the modeled and observed spectra by minimizing the chi-square:
\begin{equation}
\label{eq:delta}
\chi^2 = \displaystyle \sum_{i} \frac{ \left[\Fobs(\lami) - \Fmock'(\lami)\right]^2}{\sigma^2(\lami)},
\end{equation}
where $\Fobs$ and $\sigma$ are the observed flux and its standard deviation, \revise{$\Fmock'$} is the smoothed and rebinned $\Fmock$, and $\lami$ is the wavelength at the observational-bin center.
We adopted the observational uncertainty $\sigma$ derived in \citet{Marleau+24}, which primarily arise in the continuum subtraction procedure (see \S~2.2 in \citealp{Marleau+24}).
Our fitting targets the data points that satisfy $\Fobs>\sigma$ around the line center. 
To minimize $\chi^2$, we used \texttt{lmfit} library \citep{lmfit}.

Our fitting is scaled by two parameters scaling wavelength and flux, which correspond to $\vshift$ and $\ff\RP^2 d^{-2}$ in $\Fmock$ (see Eq.~(\ref{eq:mock_used})), respectively. By adopting the previous estimates of $d=64.5 \pm 0.4 $\,pc \citep{Bailer-Jones+2021} and $\RP=1.1\pm0.3$ \citep{Luhman.etal2023JWSTNIRSpecObservations}\footnote{Our adopted range for $\RP$ is designed to encompass all estimates and their standard deviation reported in \citet{Luhman.etal2023JWSTNIRSpecObservations} and \citet{Manjavacas.etal2024MIRSpec}}, we interpret $\ff$ as the parameter scaling the flux.
Consequently, with these two degree of freedom, the reduced chi-square can be derived as
\begin{equation}
    \redchi = \frac{\chi^2}{N-2},
\end{equation}
where $N$ is the number of spectral bins used in the fitting.

\section{Results}
\label{sec:HL_results}

\subsection{Spectral fitting}
\label{sec:fitting}
\begin{figure*}
    \centering
    \includegraphics[width=\Ftwo]{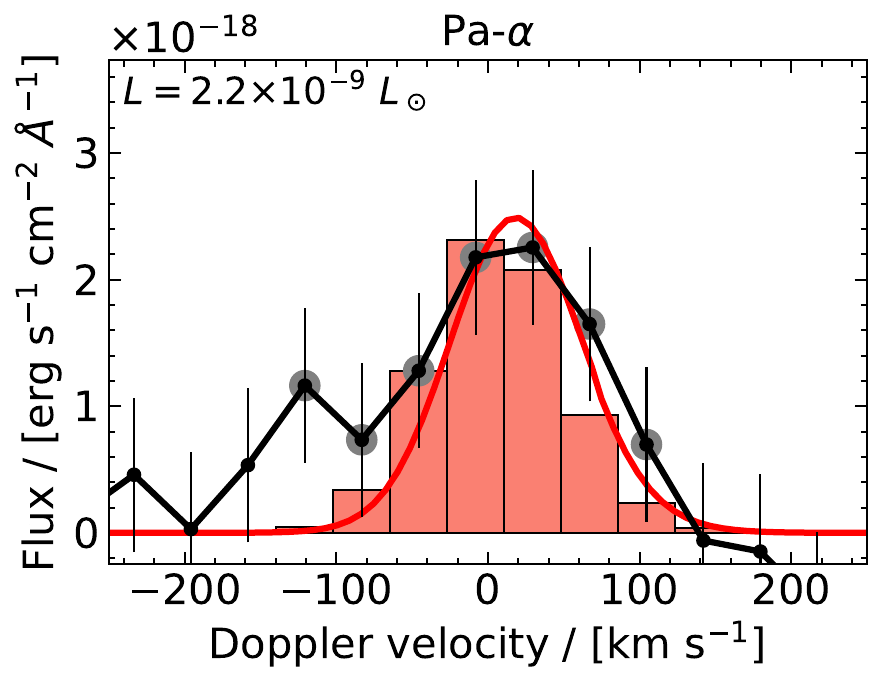}
    \includegraphics[width=\Ftwo]{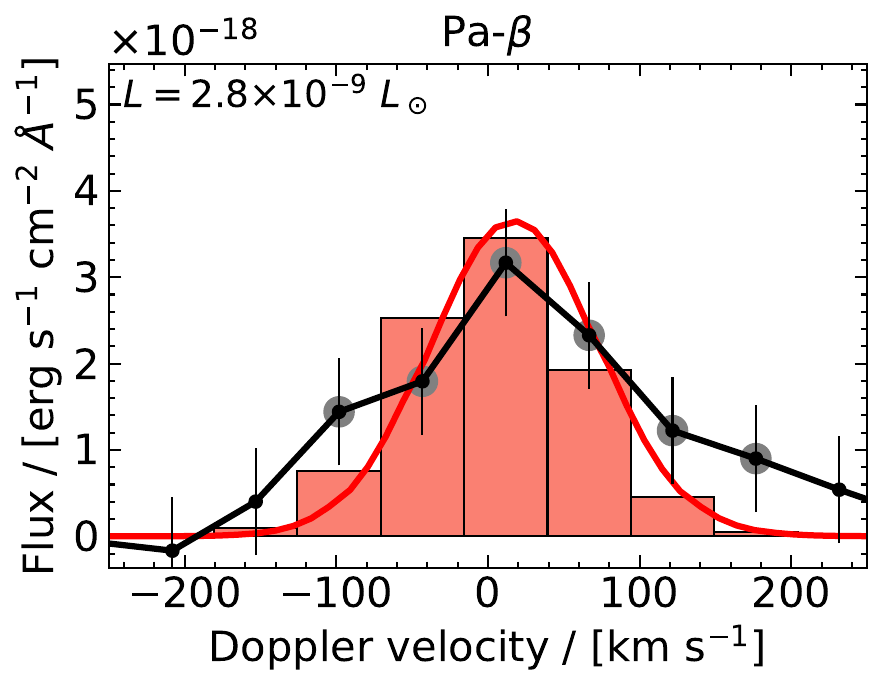}\\
    \includegraphics[width=\Ftwo]{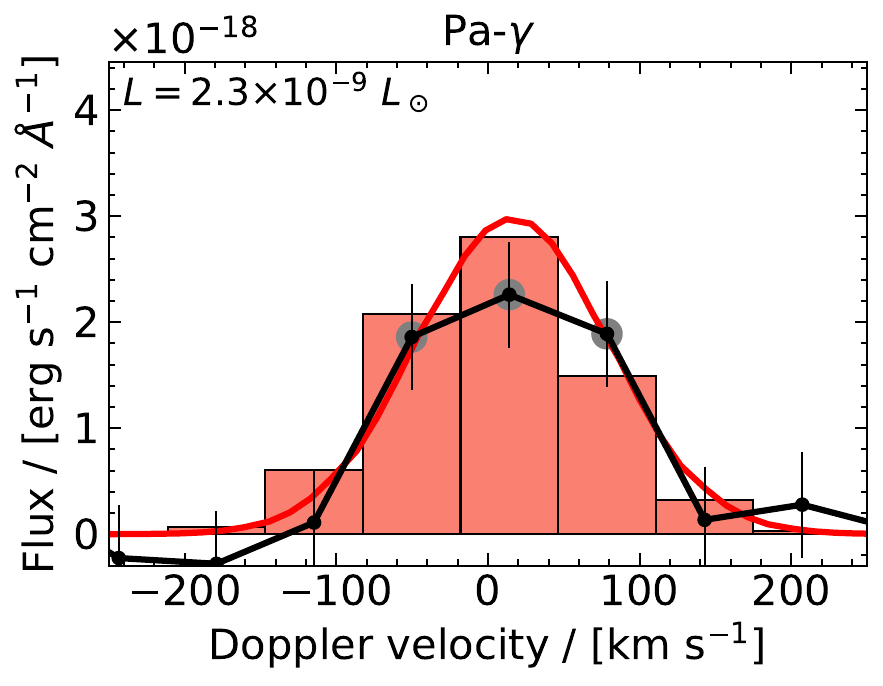}
    \includegraphics[width=\Ftwo]{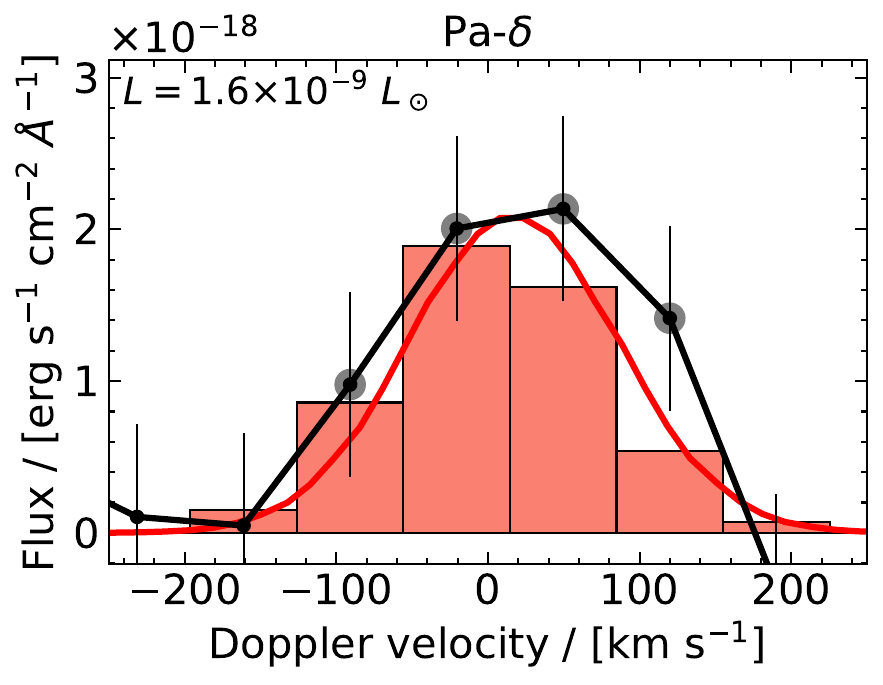}
    \caption{The modeled and observed spectra for our fiducial case of ($v_0=110\,\kms$, $n_0=10^{13}\,\cc$). The observed spectra are represented by the black with the vertical error bars. The fitting targets are denoted by larger gray circles. The red line shows the modeled spectra after accounting for instrumental broadening. The pale-red rectangles indicate the flux binned to the detector sampling.
    Line luminosities of the fitted model are labeled at upper left corner.
    }
    \label{fig:BestFit}
\end{figure*}
Firstly, we present an example of the fitted spectra. Figure~\ref{fig:BestFit} shows the spectral fitting results for our fiducial case with shock-model parameters $(v_0, n_0)=(110\,\kms, 10^{13}\,\cc)$. The rationale behind choosing these fiducial parameters is discussed subsequently. 
The black points with thin vertical error bars correspond to the observational data, and larger gray points denote the fitting targets. The red lines and pale-red histogram show the modeled spectra $\Fmock$, while both are smoothed and the latter is segmented into the observational bins (see \S~\ref{sec:M_fit}). In this fit, $\redchi=0.78$, corresponding to a $p$-value of $73\%$. 
This indicates that this fiducial case cannot be rejected at the $1-\sigma$ significance level, suggesting it is plausible within this confidence interval.

\begin{figure}
    \centering
    \includegraphics[width=\Ftwo]{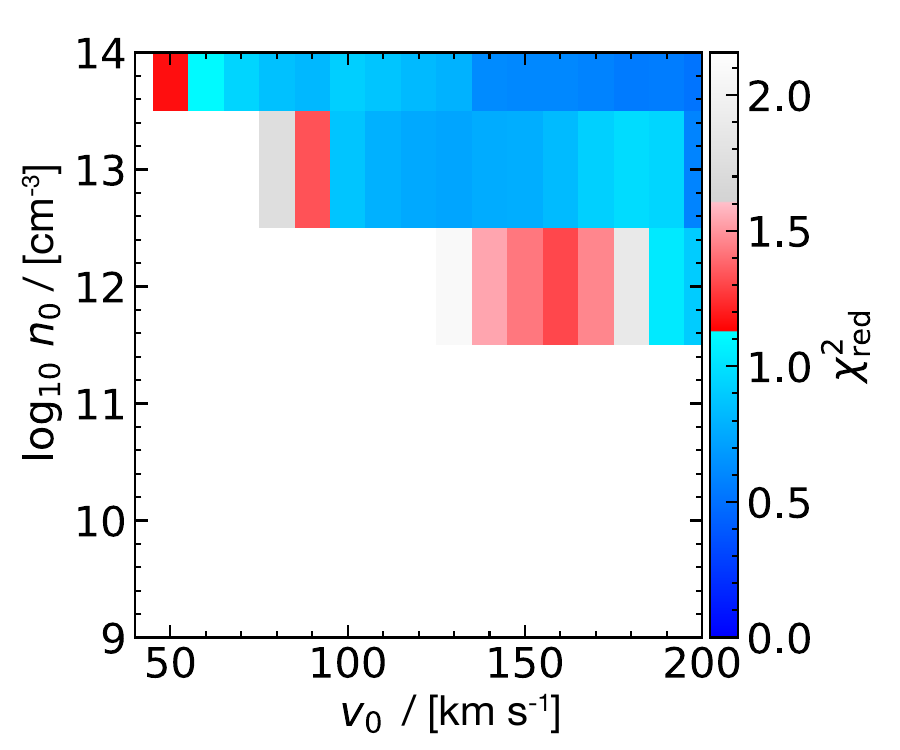}
    \caption{$\redchi$ of simultaneous fitting for \Pa, \Pb, \Pg, and \Pd of \TWAB. 
    With the degree of freedom of 18, $\redchi=1.13$ (cyan to red), $1.60$ (pink to light-gray) and $2.16$ (white) corresponds to 1-, 2-, and 3-$\sigma$ rejection of the ($v_0, n_0$).
    A denser shock is preferred. 
    }
    \label{fig:Fit_Sim}
\end{figure}

Next, we study the likelihood of a wide range of ($v_0,\,n_0$) sets. In our grid, $v_0$ ranges from~40 to~200\,\kms in steps 10\,\kms and and $n_0$ from~$10^9$ to~$10^{14}$\,\cc in steps of 1~dex. Figure~\ref{fig:Fit_Sim} shows $\redchi$ for each ($v_0,\,n_0$) set. 
The color bar highlights significant transitions: cyan to red at $\redchi=1.13$ and pink to gray at $\redchi=1.60$, marking the rejection thresholds at the 1- and 2-$\sigma$ levels, respectively, with the color bar's maximum at $\redchi=2.16$ indicating a 3-$\sigma$ rejection level.
Low densities ($n_0< 10^{13}$\,\cc) are rejected at 1-$\sigma$ (and almost $\gtrsim 3$-$\sigma$), except for high velocities ($v_0\gtrsim 200\,\kms$). Contrary to this density constraint, our fitting hardly constrains $v_0$, mainly due to the insufficient spectral resolution and sensitivity.

Within our parameter grid, the smallest $\redchi$ of $0.51$ is achieved at $(v_0, n_0)=$ $(200\,\kms,10^{14}\,\cc)$, corresponding to a $p$-value of $96\%$. However, this evaluated $v_0$ value does not align with the mass estimate deduced from photospheric-emission fittings \citep{Luhman.etal2023JWSTNIRSpecObservations,Manjavacas.etal2024MIRSpec}. 
The planetary mass and radius estimates of $\MP=5\pm2\,\MJ$ and $\RP=1.2\pm0.1\RJ$ in \citet{Luhman.etal2023JWSTNIRSpecObservations} leads to the free-fall velocity of $\vff=113\pm47\,\kms$. 
Assuming  $v_0 \approx \vff$, we adopt 110\,$\kms$ as the fiducial value for $v_0$. With this $v_0$, $\redchi$ gets the smallest at $n_0=10^{13}\,\cc$. Given that this scenario is statistically plausible within the 1-$\sigma$ level, we select (110\,$\kms$, 10$^{13}\,\cc$) as the fiducial case.
However, the $v_0$ can be smaller than $\vff$ depending on geometry, while it cannot be larger. In addition, the mass and radius estimates in \citet{Manjavacas.etal2024MIRSpec} lead to a smaller $v_0$ range of $\vff=60$--90 $\kms$, which is smaller than our fiducial $v_0$.
Furthermore, the most likely combination, while illustrative, is not definitive. Any parameter set exhibiting $\redchi<1.13$ (marked in blue) is also plausible at more than the 1-$\sigma$ significance level.
Therefore, we do not estimate the $(v_0, n_0)$ value rather only put constraints on them.

Additionally, our fitting provides estimates of fitting parameters ($\ff\RP^2d^{-2},\,\vshift$) at each ($v_0,\,n_0$). For the fiducial case of $(v_0,\,n_0)=(110\,\kms, 10^{13}\,\cc)$, these parameters are estimated to be $(1.5\pm0.1) \times10^{-24}$, and $(14\pm 4) \,\kms$, respectively.
By adopting $d=64.5 \pm 0.4 $\,pc \citep{Bailer-Jones+2021} and $\RP=1.2\pm0.1$ \citep{Luhman.etal2023JWSTNIRSpecObservations}, we convert the emitting-area ratio to the filling factor, resulting in $\ff=0.080\pm0.008\,\%$.
However, as shown in Fig.~\ref{fig:Fit_Sim}, a broad range of the model parameter is plausible. 
Considering the possible range of ($v_0$, $n_0$) within 1-$\sigma$ significance level, $\ff$ and $\vshift$ range 0.007--0.4\% and 1--24\,\kms, respectively. 
We will discuss more on this estimated $\ff$ and its implication on accretion geometry in \S~\ref{sec:geometry}.

\subsection{Estimate of mass accretion rate}
\label{sec:R_Mdot}
\begin{figure}
    \centering
    \includegraphics[width=\Ftwo]{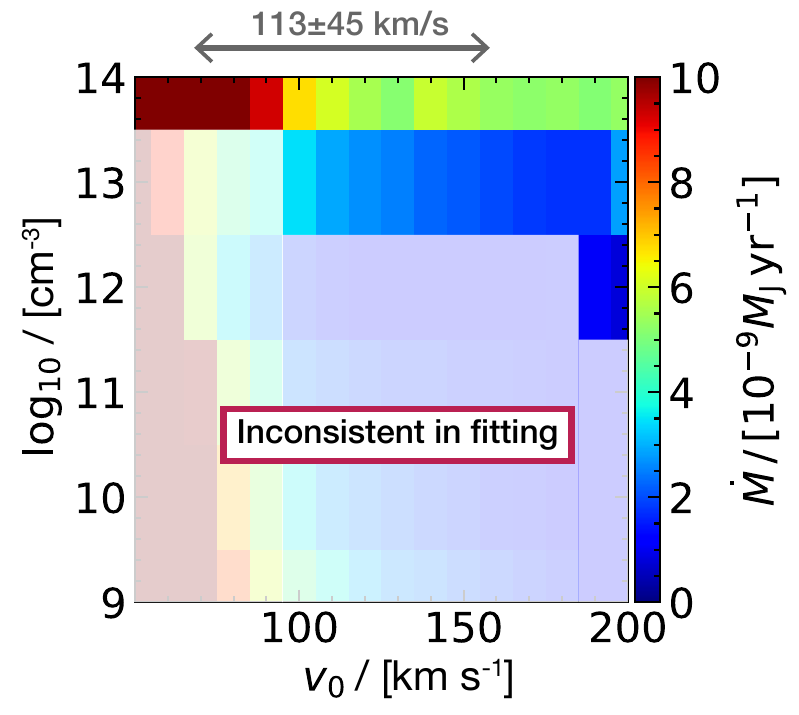}
    \caption{Mass accretion rate for each parameter set. The white shaded area indicates parameter combinations rejected at 1~$\sigma$ ($\redchi>1.13$).
    At the fiducial case of ($v_0,\,n_0$) = ($130\,\kms,\,10^{13}$\,\cc), $\Mdot=2.5\times10^{-9}\,\MJyr$.
    }
    \label{fig:Mdot}
\end{figure}

The combination of $v_0$, $n_0$ and $\ff$ gives the mass accretion rate as 
\begin{equation}
    \Mdot =  4 \pi \RP^2 \ff \mu v_0 n_0,
\end{equation}
where $\mu=2.3\times10^{-24}$\,g is the mean weight per hydrogen nuclei in the modeled abundance \citep{Aoyama.etal2018TheoreticalModelHydrogen}. Figure~\ref{fig:Mdot} shows $\Mdot$ at each ($v_0,\,n_0$), with the white shaded area denoting parameter sets excluded at the 1-$\sigma$ significance level.
For the fiducial case, $\Mdot$ is estimated to be $(2.9\pm0.2)\times10^{-9}\,\MJyr$. 
However, taking into account the range of parameters considered plausible within a 1-$\sigma$ confidence interval, $\Mdot$ ranges $1.1$--$23\times10^{-9}\,\MJyr$. 
Significantly higher $\Mdot$ values, exceeding the fiducial estimate by a factor of ten, are estimated for $v_0<90\,\kms$ and $n_0=10^{14}\,\cc$. Aside from these parameters, the variation in $\Mdot$ is relatively modest, typically within a factor of a few. Therefore, further constraints on $v_0$ may significantly narrow down this $\Mdot$ range.

The dependence of the estimated $\Mdot$ on ($v_0,\,n_0$) is due to the differential emission efficiencies of the observed lines.
A lower (greater) efficiency at a ($v_0,\,n_0$) requires more (less) $\Mdot$ to reproduce the observed flux. 
Since the postshock region radiates most of energy in \Lya\ \citep{Aoyama.etal2018TheoreticalModelHydrogen}, the \Lya flux can serve as a proxy for the total accretion energy, and thus the emission efficiency can be represented by the flux ratio of the observed Paschen lines to \Lya.
An increase in $v_0$ leads to higher temperatures in the postshock region, which elevates the hydrogen excitation level. More excitation results in higher Paschen/\Lya\ flux ratio, namely higher Paschen line emission efficiency, and thus a lower inferred $\Mdot$.
%%%
On the other hand, 
\revise{as $n_0$ increases, $\dot{M}$ decreases at lower densities ($n_0 \lesssim 10^{12}$\,\cc) and increases at higher densities.
A density increase leads to a higher optical depth in the lines}, which leads to saturation of emission due to self-absorption. The higher opacity of \Lya\ than the Paschen lines results in a more significant attenuation of \Lya\ lines, at $n_0 \lesssim 10^{12}$\,\cc.
However, at higher densities ($n_0 \gtrsim 10^{13}$\,\cc), \Lya becomes insensitive to further increase in $n_0$ because its Gaussian core profile is already fully saturated and the Lorentzian wings become dominant. In contrast, the Paschen lines continue to experience greater saturation under these conditions.
Therefore, with $n_0$ increasing, the emission efficiency of Paschen lines initially increases and then decreases, corresponding to a decrease and then an increase of estimated $\Mdot$ at lower and higher $n_0$, respectively.

Our $\Mdot$ estimate naturally agrees with the $\Mdot\approx 5\times 10^{-9}\,\MJyr$ that \citet{Marleau+24} estimated, since it depends on the same shock-emission model, while they used only the total flux of each line in that work. Compared with the estimates from empirical relationships obtained for Young Stellar Objects (YSOs; e.g., \citealp{Alcala.etal2017Xshooterspectroscopyyoung}), our estimate is higher \citep[see][]{Marleau+24}. This is because in the planetary case, the optically thicker postshock region radiates more energy through the non-observed Lyman lines in Lorentzian broad wings as discussed in \citet{Aoyama.etal2021ComparisonPlanetaryHaemission}.

\section{Discussion}
\label{sec:HL_Discussion}

\subsection{Constraints on accretion geometry}
\label{sec:geometry}
The $\ff$ estimated in \S~\ref{sec:fitting} serves as an indicator of the accretion geometry. An important open question regarding the planetary accretion geometry is whether CPD is connected to the planetary surface or not. The CPD become disconnected if magnetospheric accretion happens, that is, if the magnetic field of the planet is strong enough and connected well enough to the gas to disrupt the CPD \citep{Zhu.etal2015STRUCTURESPIRALSHOCKS,Batygin2018TerminalRotationRates,Hasegawa+2021,Hasegawa+2024}. In this scenario, the accretion flow concentrates along the object's magnetic field, leading to a small $\ff$ \citep[e.g.,][]{Valenti.etal1993Tauristarsblue}. Our estimated small $\ff$ of 0.007--0.4\% aligns with this scenario and is consistent with typical values of 0.01--1\% at very-low-mass objects \citep{Herczeg.Hillenbrand2008UVExcessMeasures}, and is smaller than the 0.1--10\% observed in higher mass starts \citep{Valenti.etal1993Tauristarsblue,Ingleby.etal2013ACCRETIONRATESTAURI}, following the trend of decreasing $\ff$ from higher to lower mass stars.

Conversely, if the CPD is connected to the planet, planetary-surface accretion shocks capable of producing significant hydrogen line emission result either from the direct accretion from the protoplanetary disk or from the fast accretion layer on the CPD surface \citep{Takasao.etal2021HydrodynamicModelHa}.
Given that direct accretion from the protoplanetary disk falls almost uniformly on the planetary surface \citep{Marleau+2023}, \ff\ is on the order of $100\%$, which is inconsistent with this observation. 
Also for the fast accretion layer scenario, while its accretion shock is localized, its $\ff$ is order of $10\%$, significantly larger than our estimate \citep{Takasao.etal2021HydrodynamicModelHa}.
On the other hand, through their use of equilibrium-opacity tables, \citet{Marleau+2023} showed that efficient cooling following the CPD surface shock could result in a much thinner fast accretion layer, potentially aligning the \ff\ with our estimates. However, the flux from such a thin layer is negligible compared to that from direct accretion from the disk to the planetary surface. Therefore, the scenario of a CPD connected to the planet appears inconsistent with our observations.

Another possible source of hydrogen line emission is the shock on the CPD surface \citep[e.g.,][]{Aoyama.etal2018TheoreticalModelHydrogen}. 
Given the critical velocity for emitting hydrogen lines of $\vcrit \approx 30\,\kms$ \citep{Aoyama.etal2018TheoreticalModelHydrogen}, the critical radius where the free-fall velocity exceeds $\vcrit$, is 
\begin{eqnarray}
    \rcrit &=& 2G\MP \vcrit^{-2} \\
    &\approx& 20 \RJ
    \left( \frac{\MP}{5\MJ} \right) 
    \left( \frac{\vcrit}{30\kms} \right)^{-2}.\displaystyle
    \label{eq:rcrit_value}
\end{eqnarray}
The area inside $\rcrit$ well represents the emitting area because the emission from the weaker shock, namely outer and closer to $\rcrit$, contributes more to the flux compare to the stronger shock closer to the planet, mainly due to its larger annular area \citep{Aoyama.etal2018TheoreticalModelHydrogen}.
Considering that the emission originates from both sides of the CPD, the filling factor, which is the emitting area normalized by the planet surface area, can be estimated as \citep{Aoyama.etal2018TheoreticalModelHydrogen}
\begin{eqnarray}
\displaystyle
    \ff &=&  \frac{2 \pi \rcrit^2}{4 \pi \RP^2} \\  %
    &\approx& 50
    \left( \frac{\MP}{5\MJ} \right)^2 
    \left( \frac{\RP}{2\RJ} \right)^{-2}
    \left( \frac{v_\mathrm{crit}}{ 30\,\kms} \right)^{-4}.
\end{eqnarray}
Except for extreme cases such as highly concentrated shock-generating flow, $\ff$ should be much larger than we estimated for \TWAB if the CPD-surface emission dominated. In addition, although \citet{Marleau+2023} demonstrated that $\rcrit$ can be around half of the simple estimate of Eq.~(\ref{eq:rcrit_value}), mainly due to the slopes of accretion flow and CPD surface (see the right panel of their Fig.~2), the emitting area is still larger than the planetary surface area.
Note that the CPD surface shock itself is not ruled out because, when both the CPD and planetary surface shocks emit hydrogen lines, the planetary surface shock dominates the total flux \citep{Aoyama+2020}. The higher density and velocity more than compensate for the smaller emitting area \citep{Marleau+2023}.

In summary, our analysis suggest that magnetospheric accretion is occurring at \TWAB.

\subsection{Possible emission from preshock accretion flow}
\label{sec:EMechanism}
The emission mechanism of hydrogen lines at planetary masses remains a subject of debate. This work focuses on emission from the postshock gas, which always emits hydrogen lines in planetary-mass objects because the hydrogen lines are much efficient coolant than the others at the planetary postshock temperature \citep{Aoyama.etal2018TheoreticalModelHydrogen}. However, in the magnetospheric accretion scenario, the preshock accretion flow can emit hydrogen lines \citep{Thanathibodee.etal2019MagnetosphericAccretionSource}, and may even dominate postshock emission in highly accreting massive objects \citep{Aoyama.etal2021ComparisonPlanetaryHaemission}. 
Distinguishing between these emission mechanisms is essential for properly interpreting hydrogen line observations.

The only observational confirmation of shock-heated emission (or absence of strong preshock emission) was reported for \Delormeb by \citet{Betti.etal2022NearinfraredAccretionSignatures,betti22c}. They compared the hydrogen-line emission color (flux ratios of multiple hydrogen lines) with both the postshock emission model and the preshock accretion-flow-emission model, finding better consistency with the former. 
This subsection applies a similar emission-color diagnostic to validate our assumption that postshock emission dominates the detected hydrogen lines in \TWAB.

\subsubsection{Methodology of emission-color diagnostic}
Following \citet{Betti.etal2022NearinfraredAccretionSignatures,betti22c}, we use the modeling frameworks of \citet{Aoyama.etal2018TheoreticalModelHydrogen} for postshock emission and
\citet{Kwan.Fischer2011OriginsHeCa} for preshock emission. 
While we use the postshock emission model identically as in \citeauthor{betti22c}, we introduce the following differences for the preshock emission model:
\begin{itemize}
    \item We use different collisional electron-transition coefficients from the ones in the shock model \citep{Aoyama.etal2018TheoreticalModelHydrogen}. Specifically, the coefficients for higher states (principal quantum number $\np>5$) are given by \citet{Vriens.Smeets1980Crosssectionrateformulas} rather than extrapolating the fits from the coefficients for low levels ($\np \leq 5$) \citep{Anderson.etal2002Rmatrixpseudostatesapproach}.
    \item The ionization and electron transitions of helium and metals are not solved. Instead, we estimate their electron supply using the Saha equation \citep[e.g.,][]{spitzer1998physicalprocessesinterstellar}.
    \item As the photospheric irradiation from the accreting objects, we adopt an atmospheric model with the effective temperature $T_*=4000\,K$, solar metallicity, and the surface gravity $\log g = 3$ \citep{Catelli+Kurucz2003}\footnote{The data is from \citet{STScI} and is available at \url{https://archive.stsci.edu/hlsps/reference-atlases/cdbs/grid/k93models}.}, which we expect to correspond to ``a CTTS of temperature $T_*=4000$\,K'' as used in \citet{Kwan.Fischer2011OriginsHeCa}. 
    We tested the weaker irradiation from a planetary mass object and found no significant difference even when replacing the photospheric irradiation by $3000$\,K blackbody radiation.
\end{itemize}
Although these differences affect the emission color correlation to model parameters, they hardly change the scatter plot distributions in Fig.~\ref{fig:ratio_comp}. Therefore, our model comparison works similarly to the one in \citet{Betti.etal2022NearinfraredAccretionSignatures,betti22c}. 

\subsubsection{Hydrogen-line emission color at TWA 27 B}
\begin{figure*}
    \centering
    \includegraphics[width=\textwidth]{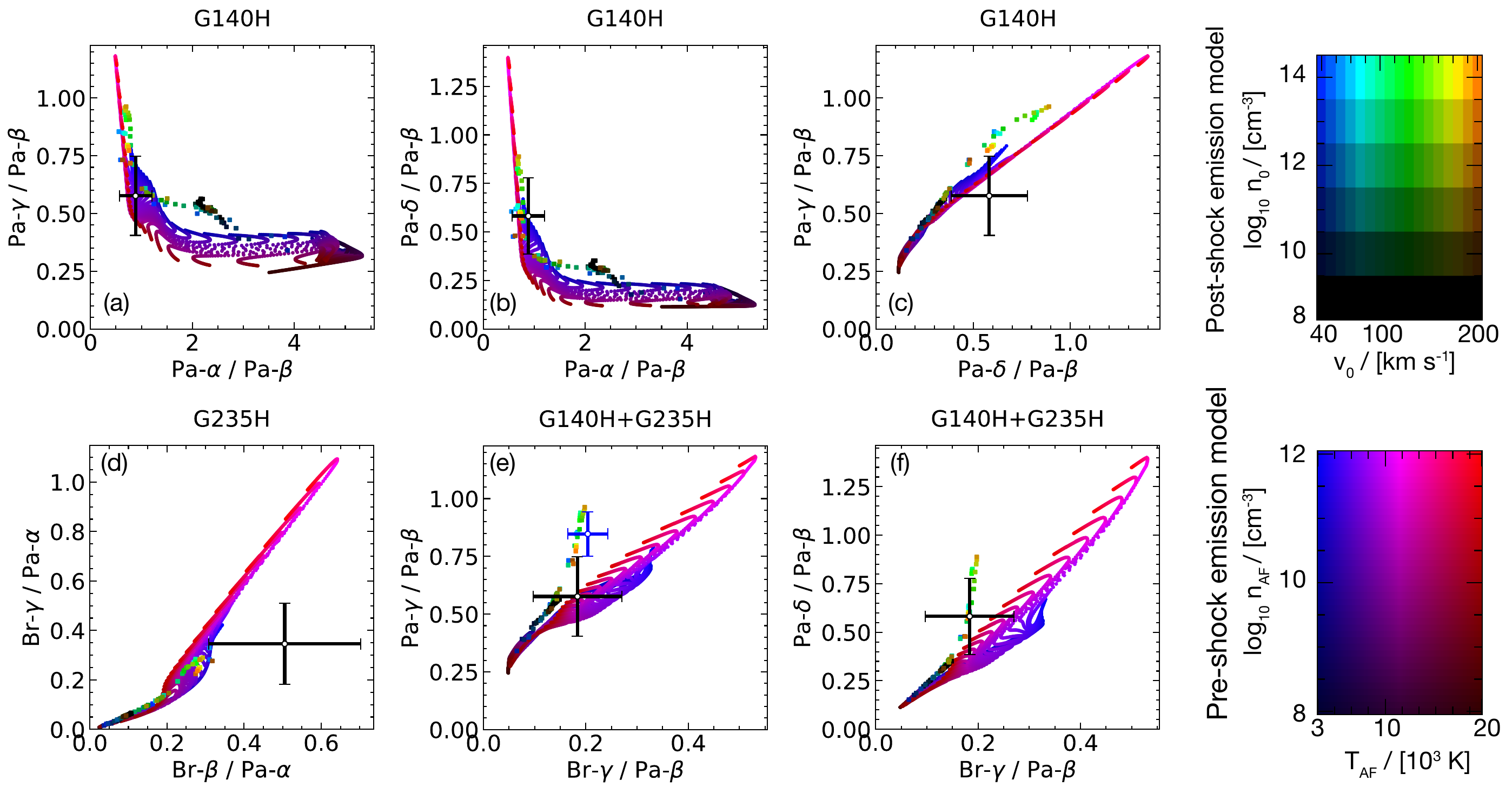}
    \caption{
    Flux ratios or emission color of several hydrogen lines. The blue--orange squares represent the predictions for postshock emission, while the blue--red points correspond to the preshock emission. Observational results are marked by a black circle with error bars.
    Across all panels, the observed emission color aligns with both emission models due to large uncertainty.
    \revise{As a reference, the successful case of \Delormeb, analyzed in \citet{betti22c}, is shown as blue cross in panel~(e).
    }
    }
    \label{fig:ratio_comp}
\end{figure*}

Figure~\ref{fig:ratio_comp} illustrates the emission colors corresponding to both postshock and preshock emission mechanisms, compared to the observed data. The observed emission color is shown as black circles with error bars. For postshock emission, colors vary as a function of shock velocity $v_0$ (blue to orange) and number density $n_0$ (darker to brighter), while for preshock emission, they vary with accretion flow temperature $\TAF$ (blue to red) and number density $\nAF$ (darker to brighter).
Panels~(a), (b), and~(c) display the emission color using \Pa, \Pb, \Pg, and \Pd\ fluxes from the G140H filter. Similarly, panel (d) shows the emission color among \Pa, \Brb, and \Brg\ from the G235H filter. Additionally, panels (e) and (f) show the emission color from combination of different filters. 
Given that observations using both filters were conducted only around 40-min apart, no time variation is anticipated, as supported by the nearly consistent \Pa\ fluxes \citep{Marleau+24}.
All the observed line ratios that we use are listed in table~\ref{tab:ratios}. In all the panels, the observed emission color is consistent with both post- and preshock emission models, mainly due to large observational uncertainties.

Despite the current inability to distinguish conclusively between the emission mechanisms for \TWAB, this observation provide valuable insights for guiding future observations. It is important to note that even with higher precision, this diagnostic approach may not be effective if the observed emission-color is reproducible with both the emission mechanisms.
Such indistinguishable emission-color is indicated by the overlap of predicted emission-color from the two models in Figure~\ref{fig:ratio_comp}. 
The combination of \Pa-\Brb-\Brg\ in panel~(d) is generally ineffective, as the emission-color distributions for the two mechanisms nearly coincide. 
While \Pa-\Pb-\Pg\ (a) and \Pa-\Pb-\Pd\ (b) could potentially be useful, they are not suitable for \TWAB\ due to overlapping distributions near the observed emission color.
In contrast, \Pb-\Pg-\Pd (c), \Pb-\Pg-\Brg (e) and \Pb-\Pd-\Brg (f) demonstrate the potential to differentiate between the mechanisms with more precise observations.
Indeed, \citet{Betti.etal2022NearinfraredAccretionSignatures,betti22c} successfully distinguished between the mechanisms with \Pb-\Pg-\Brg\ (\revise{blue cross in} our panel [e]), benefiting from a signal-to-noise ratio that was twice as good as ours. Furthermore, for the \Pb-\Pg-\Brg\ (panel [e]) and \Pb-\Pd-\Brg\ (panel [f]), higher values of $v_0$, $n_0$, or $\nAF$ result in greater deviations between two emission models. The larger mass of their target (\Delormeb; 12--14$\MJ$ \citealp{Eriksson.etal2020Strongemissionsigns}) also plays a role in their success.

\begin{table}
\centering
    \caption{Observed flux ratios, grouped by grating.}
    \begin{tabular}{c|c}
    \hline
    \multicolumn{2}{c}{G140H}\\
    \Pa/\Pb  &  $0.88\pm0.32$ \\
    \Pg/\Pb  &  $0.58\pm0.17$ \\
    \Pd/\Pb  &  $0.58\pm0.20$ \\ \hline
    \multicolumn{2}{c}{G235H}\\
    \Brb/\Pa  &  $0.51\pm0.19$ \\
    \Brg/\Pa  &  $0.35\pm0.17$ \\ \hline
    \multicolumn{2}{c}{G140H+G235H}\\
    $\Brg_\mathrm{G235H}/\Pb_\mathrm{G140H}$ & $0.18\pm0.08$\\
    \hline
    \end{tabular}
    \tablecomments{The original data were reported by \citet{Luhman.etal2023JWSTNIRSpecObservations}, and the data with continuum subtraction were presented by \citet{Marleau+24}.}
    \label{tab:ratios}
\end{table}

\subsection{Future observations for better constraints on accretion properties}
\label{sec:HL_Discussion_SimFit}

While we have estimated accretion properties and geometry, it is accompanied by substantial uncertainties. 
This subsection delves into strategies for achieving more precise constraints in future observations. Specifically, we will discuss how better observational quality enhances the analysis of spectral profiles and hydrogen line emission colors and the resulting constraints.

A higher spectral resolution offers access to more detailed and accurate spectral profiles, which can be smoothed out in low-resolution data.
Notably, the spectral profile (width) of Paschen lines is more sensitive to $v_0$ than $n_0$ \citep{Aoyama.etal2018TheoreticalModelHydrogen,Aoyama.etal2021ComparisonPlanetaryHaemission}. 
With the current spectral resolution of $R \approx 2000$, we can only weakly reject low velocities ($v_0<90\,\kms$) at a significance level of about 1-$\sigma$, and provide limited constraints at higher $v_0$ values (see also Appendix~\ref{sec:A_fitting}). 
Additionally, a higher signal-to-noise ratio (SNR) would also improve the spectral fitting analysis, because the fainter part of the lines has a wider profile (the ``wings'' of the line profile). Notably, \citet{Demars+2023} derived a more conclusive outcome from spectral profile even with a similar spectral resolution of $R\approx2500$ but with a peak $\textrm{SNR} \approx 10$ instead of only $\approx 4$ as we have. Enhancing spectral profile quality would enable better constraints on $v_0$, potentially allowing for an independent estimate of $\MP$ and $\RP$ based solely on hydrogen line observation, separate from estimate derived from the photospheric-emission observations \citep{Luhman.etal2023JWSTNIRSpecObservations,Manjavacas.etal2024MIRSpec}.

On the other hand, emission color analysis primarily relies on the precision of flux measurements, provided the spectral resolution is sufficient to distinguish or remove other lines and the (pseudo-)continuum. This analysis is crucial not only for exploring emission mechanisms, as discussed in Section~\ref{sec:EMechanism}, but also for constraining accretion properties. Notably, the rejection of low $n_0$ values and subsequent constraints on accretion geometry largely stem from emission color as well, as detailed in Appendix~\ref{sec:ratio}. For a better understanding of accretion dynamics and associated hydrogen line emission, precision of flux measurement is therefore indispensable.

In summary, a higher-resolution spectral profile and a greater precision in flux measurements lead to better constraints in $v_0$ and $n_0$, respectively. While enhancing both would be ideal, we suggest to prioritize the flux precision, which is a more important factor to elucidate the accretion geometry and emission mechanisms.

\section{Summary and conclusion}
\label{sec:HL_summary}
We have analyzed the hydrogen lines at \TWAB\ observed with \NIRSpec\ \citep{Luhman.etal2023JWSTNIRSpecObservations} and continuum-subtracted \citep{Marleau+24}, using the accretion-shock emission model of \citet{Aoyama.etal2018TheoreticalModelHydrogen}.
We have fitted the simultaneously-observed four bright Paschen lines of \Pa, \Pb, \Pg, and \Pd\ with the modeled spectra and put constraints on the accretion at \TWAB. Our major findings are as follows:
\begin{enumerate}
    \item The shock-emission model reproduces well the observed Paschen lines when the accretion gas immediately before the shock has a high density of $n_0 \gtrsim 10^{13}$\,\cc. Lower densities are rejected with a significance of at least 1-$\sigma$ and mostly, depending on the preshock velocity $v_0$, $>3$-$\sigma$.
    \item
    The mass accretion rate is estimated to be $\Mdot = (2.9\pm 0.2)\times 10^{-9}$\,\MJyr\ for our fiducial case characterized by the shock parameters ($v_0,\,n_0$)=($130\,\kms,\,10^{13}\,\cc$). However, considering the parameter range within a 1-$\sigma$ confidence interval, the uncertainty in $\Mdot$ extends to approximately an order of magnitude.
    Our estimate is naturally consistent with the previous estimate \citep{Marleau+24} that applied to the observed, continuum-subtracted, wavelength-integrated fluxes line-luminosity--accretion-luminosity correlations based on the same shock-emission model \citep{Aoyama.etal2021ComparisonPlanetaryHaemission}.
    \item Combined with the planetary mass and radius estimated from the photospheric-emission fitting \citep{Luhman.etal2023JWSTNIRSpecObservations,Manjavacas.etal2024MIRSpec}, our estimated preshock density implies that the accretion filling factor on planetary surface is $\ff\approx0.007$--0.4\%.
    Such a small filling factor can be achieved in magnetospheric accretion \citep[e.g.,][]{Hartmann.etal2016AccretionPreMainSequenceStars} but is inconsistent with the accretion from the protoplanetary disk to the planet \citep{Marleau+2023}, CPD-surface fast accretion to the planet \citep{Takasao.etal2021HydrodynamicModelHa}, and the CPD surface shock \citep{Aoyama.etal2018TheoreticalModelHydrogen}. 
    The hydrogen line emission from the CPD surface shock is not ruled out but has at most a minimal contribution to the observed flux relative to the dominant emission from the magnetospheric accretion.
    Also, this estimate is largely consistent with observations of very-low-mass-stars, for which $\ff\sim0.01$--$1\%$ \citep{Herczeg.Hillenbrand2008UVExcessMeasures}.
    \item We have attempted to examine the contribution of emission from the preshock accretion flow \citep[e.g.,][]{Thanathibodee.etal2019MagnetosphericAccretionSource} with the emission color (flux ratios) of hydrogen lines. However, the emission color is consistent with the emission from both the preshock and postshock emission models, mainly due to the large observational flux uncertainties. Therefore, it is not ruled out that the preshock accretion flow can contribute to and even be dominant in the observed flux.
    Within the hydrogen lines accessible with \NIRSpec, a combination of three lines from \Pb, \Pg, \Pd, and \Brg\ proves suitable for distinguishing between them, with greater precision (see panels e and f in Fig.~\ref{fig:ratio_comp}).
\end{enumerate}

While \NIRSpec\ observation provides valuable constraints on the mass accretion rate and accretion geometry, follow-up observations with higher spectral resolution and/or deeper sensitivity remains crucial for reducing the current uncertainties. In particular, to confirm the mechanism of hydrogen-line emission, higher precision in flux measurements of multiple hydrogen lines is essential.

\section*{Acknowledgments}
Y.A.\ acknowledges support by the China Postdoctoral Science Foundation under grant No.~2023M740110, and the National Science Foundation of China with grant No.~12233004.
G.-D.M.\ acknowledges the support of the DFG priority program SPP 1992 ``Exploring the Diversity of Extrasolar Planets'' (MA~9185/1) 
and from the European Research Council under the Horizon 2020 Framework Program via the ERC Advanced Grant ``Origins'' (PI: Henning), Nr.~832428.
% G-DM also acknowledges the support from the Swiss National Science Foundation under grant
% 200021\_204847
% ``PlanetsInTime''.
% Parts of this work have been carried out within the framework of the NCCR PlanetS supported by the Swiss National Science Foundation.
%
The JWST data presented in this article were obtained from the Mikulski Archive for Space Telescopes (MAST) at the Space Telescope Science Institute. The specific observations analyzed can be accessed via \dataset[10.17909/2f9b-ea80]{https://doi.org/10.17909/2f9b-ea80}.

\software{
astropy \citep{astropy:2013,astropy:2018,astropy:2022}, 
lmfit \citep{lmfit},
Matplotlib (\citet{Hunter2007}),
Numpy (\citet{Harris+2020}),
Python3 (\citet{Python3}),
scipy \citep{2020SciPy-NMeth},
SUNDIALS/CVODE \citep{SUNDIALS2005,SUNDIALS2022,CVODE}
}

\appendix

\section{Spectral fitting on each line}
\label{sec:A_fitting}
\begin{figure*}
    \centering
    \includegraphics[width=\Ftwo]{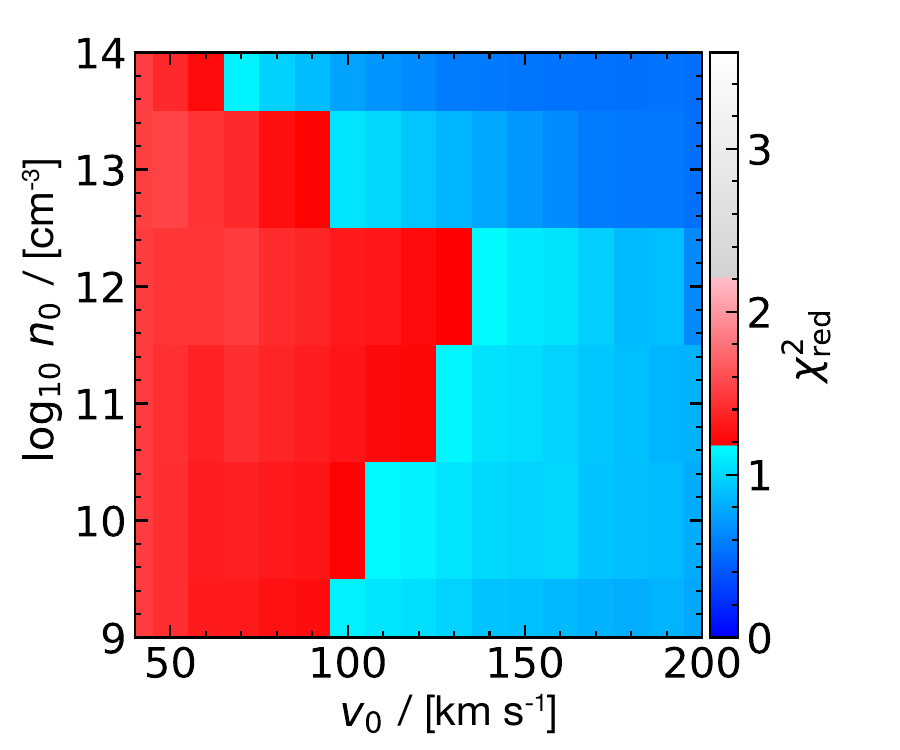}
    \includegraphics[width=\Ftwo]{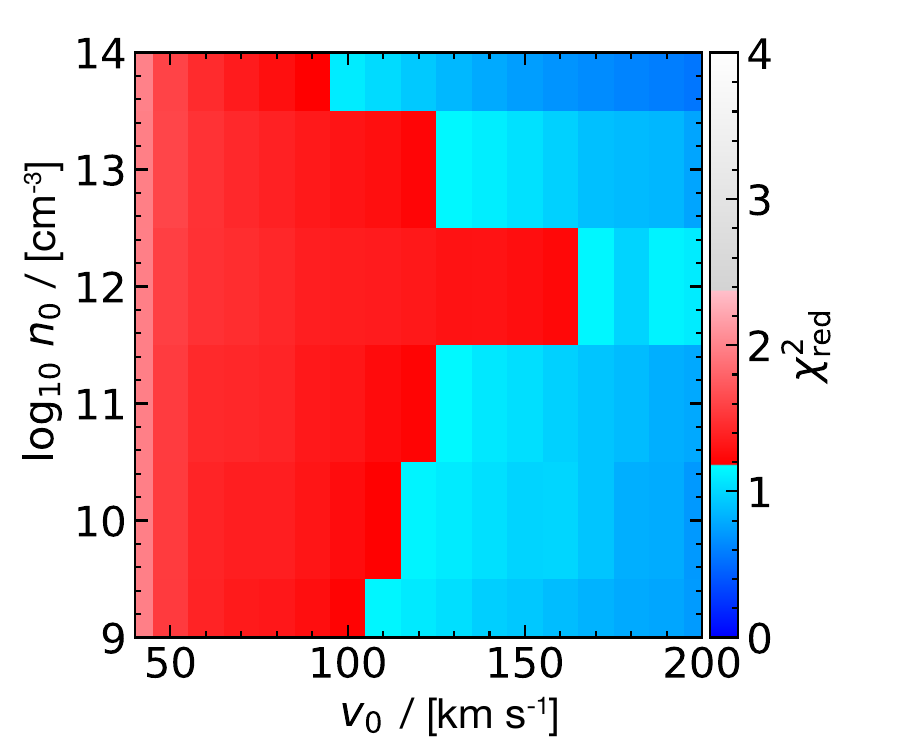}
    \includegraphics[width=\Ftwo]{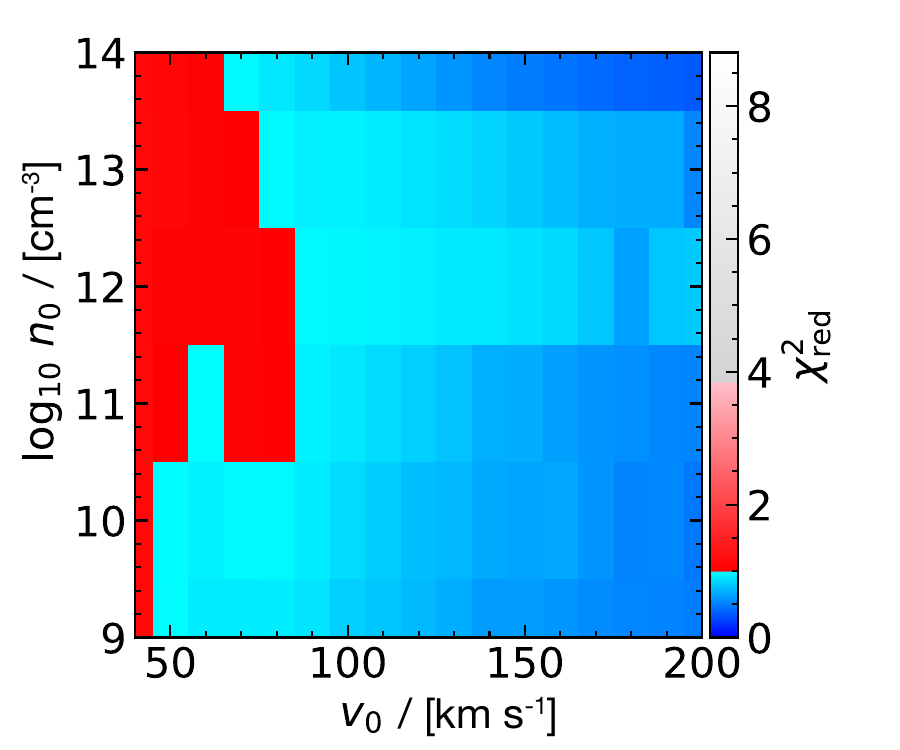}
    \includegraphics[width=\Ftwo]{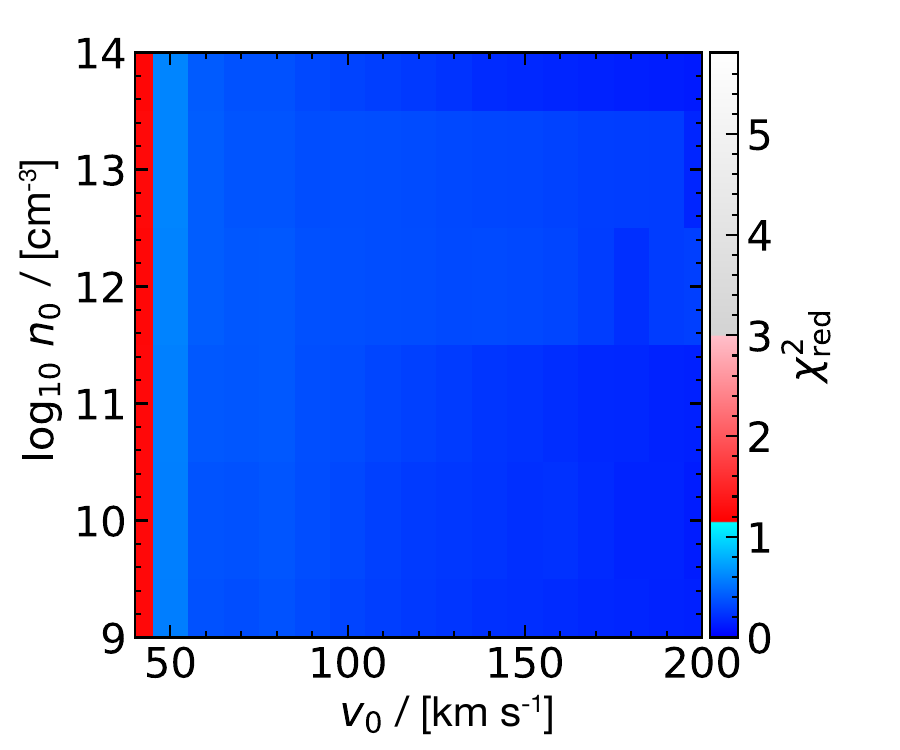}
    \caption{Over our model grid, $\redchi$ of fitting to \Pa (upper left), \Pb (upper right), \Pg (lower left), and \Pd\ (lower right) of \TWAB. Similar to Figure~\ref{fig:Fit_Sim} but independently fitting to each line. 
    \revise{The degrees of freedom are 5, 4, 1, and 2 for \Pa, \Pb, \Pg, and \Pd, respectively.}
}
    \label{fig:chi2_each}
\end{figure*}

Figure~\ref{fig:chi2_each} presents $\redchi$ for \Pa, \Pb, \Pg, and \Pd, similarly to Fig.~\ref{fig:Fit_Sim}. Note that the different number of degrees of freedom changes the correspondence between $\redchi$ and $p$-value. The sharp color changes of cyan--red and pink--gray are where the $p$-value corresponds to 1- and 2-$\sigma$, and maximum in color bar corresponds to 3-$\sigma$.
Differently from the case in Figure~\ref{fig:Fit_Sim}, each line here is fitted independently, making them unaffected by relative flux consideration, as discussed in \S~\ref{sec:HL_Discussion_SimFit}.
Due to the low spectral resolution and signal-to-noise ratio, the observed and modeled spectra match within a 3-$\sigma$ confidence interval. However, at 1-$\sigma$ significance level, a higher velocity is preferred.

\section{Flux ratio analysis}
\label{sec:ratio}
\begin{figure}
    \centering
    \includegraphics[width=\Ftwo]{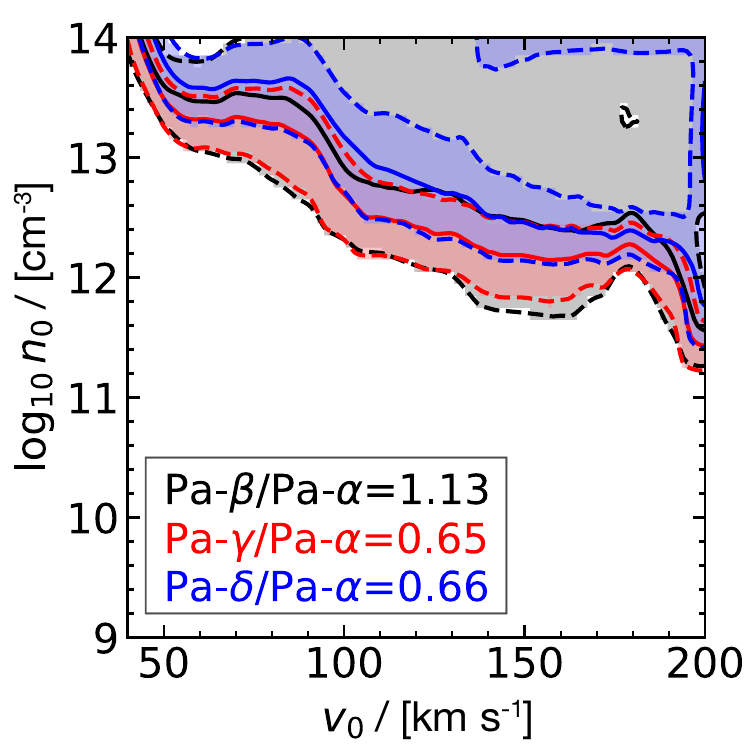}
    \caption{Constraint of shock parameters from flux ratios. The solid lines represent the parameter sets where our model reproduces the observed flux ratios of \Pb/\Pa\ (black), \Pg/\Pa\ (red), and \Pd/\Pa\ (blue). The shaded region, bordered by the dashed lines, encompass values that agree with the observation within 1-$\sigma$ uncertainty. 
    }
    \label{fig:ratio}
\end{figure}

Figure~\ref{fig:ratio} shows the shock parameters $(v_0,\,n_0)$ where our model reproduces the observed flux ratios of four Paschen lines. For the three flux ratios derived from the four lines, the solid contour indicates the parameters where our model aligns with the mean ratio of observed fluxes. The shaded areas, bordered by the dashed lines, represent the standard deviation. The wavelength-integrated fluxes and their uncertainties are listed in \citet{Marleau+24}.
All three-colored shades overlap between the lower blue and the upper red dashed lines, around $n_0\sim10^{13}$\,\cc. 
The distribution of these shades roughly correspond to the $\redchi<1.13$ region in Figure~\ref{fig:Fit_Sim}, as expected.

Flux ratios of hydrogen lines with different wavelengths can be significantly affected by extinction. Consequently, these flux ratios can constrain the extinction degree. Such an approach was previously attempted for \PDSbc by \citet{Hashimoto.etal2020AccretionPropertiesPDS} and \citet{Uyama.etal2021KeckOSIRISPav} who used the ratio of non-detection upper limit of \Hb\ and \Pb\ to the detected flux of \Ha. 
However, for Paschen lines with $\lambda>1 \micron$, extinction has a much smaller impact on the flux ratios. In this study, using only Paschen and Brackett lines, we confirm that a visible extinction degree of $\AV<5$\,mag hardly changes the flux ratios, based on the extinction curve of interstellar dust \citep{Wang.Chen2019OpticalMidinfraredExtinction}. 
To effectively evaluate the extinction degree from the flux ratio analysis, lines at short wavelengths, such as Balmer and Lyman lines, should be considered.

\bibliographystyle{aasjournal}
\bibliography{new}

\end{document}